# Spin torque control of antiferromagnetic moments in NiO

Takahiro Moriyama[1,2], Kent Oda[1], and Teruo Ono[1,2]

[1] *Institute for Chemical Research, Kyoto University, Uji, Kyoto, 611-0011, Japan*
[2] *Center for Spintronics Research Network, Osaka University, Toyonaka, Osaka, 560-8531, Japan*

**For a long time, there have been no efficient ways of controlling antiferromagnets. Quite a strong magnetic field was required to manipulate the magnetic moments because of a high molecular field and a small magnetic susceptibility[1, 2, 3]. It was also difficult to detect the orientation of the magnetic moments since the net magnetic moment is effectively zero. For these reasons, research on antiferromagnets has not been progressed as drastically as that on ferromagnets which are the main materials in modern spintronic devices. Here we show that the magnetic moments in NiO, a typical natural antiferromagnet, can indeed be controlled by the spin torque with a relatively small electric current density (~5 x $10^7$ A/cm$^2$) and their orientation is detected by the transverse resistance resulting from the spin Hall magnetoresistanc[4]. The demonstrated techniques of controlling and detecting antiferromagnets would outstandingly promote the methodologies in the recently emerged "antiferromagnetic spintronics" [5, 6]. Furthermore, our results essentially lead to a spin torque antiferromagnetic memory.**

The majority of spintronics research and applications has so far dealt with ferromagnetism, with much less attention given to antiferromagnetic materials. Although they have no net magnetic moment, the microscopic magnetic moments in antiferromagnetic materials can in principle exhibit a similar spintronic effect, such as various magnetoresistance effects and the spin torque effect[7], as seen in ferromagnetic materials[8].

Recent theoretical and experimental studies have suggested that it is possible to control the antiferromagnetic moments by the spin torque due to a consequence of the interaction between the local moment and the itinerant electron spins[9, 10, 11, 12, 13, 14]. It is also predicted and partly demonstrated that magnetoresistances, such as anisotropic magnetoresistance[15, 16], the planer Hall effect[17], and the spin Hall magnetoresistance[18], were available for detecting the orientation of the magnetic moments in antiferromagnets.

Wadley et al.[14] have recently reported that the tetragonal phase CuMnAs having a broken inversion symmetry in its spin sublattices gives rise to the spin torque by a flow of an electric current in itself. Although their seminal reports[13, 14, 19] seem to magnificently advance a spin torque operation of the antiferromagnet, choice of the materials having such a complex unit cell structure is quite limited. To further pave a wide pathway of antiferromagnetic spintronics, it is desirable to seek the spin torque control of more general antiferromagnetic materials.

In this letter, we show the spin torque control of the orientation of the magnetic moments in an antiferromagnetic NiO and also show an electrical detection of the orientation of the magnetic moments. Our results essentially demonstrate a spin torque antiferromagnetic memory with NiO.

NiO is one of the most common natural antiferromagnetic oxides the study of which dates back to the dawn of the antiferromagnetism[20]. It still continues to be an archetype material for investigating interesting novel phenomena, such as the THz magnon dynamics[21], the spin current transmission[22, 23], and the spin Hall magnetoresistance[18]. NiO has a rock-salt structure with magnetic moments inhabiting on Ni cations. The magnetic moments lie in (111) planes and are ferromagnetically aligned on an {111} plane and they antiferromagnetically couples with the magnetic moments in neighboring {111} planes by super exchange coupling[20].

Layer structures employed in this study are Pt 4 nm/ NiO $t_{NiO}$ nm/ Pt 4 nm ($t_{NiO}$ = 0 ~ 100 nm) and Pt 4 nm/ SiO 10 nm/ Pt 4 nm formed on a thermally oxidized Si wafer by magnetron sputtering (see Methods for details.). Figure 1 (a) shows the basic principle of the spin torque rotation of the antiferromagnetic moments in a Pt/ NiO/ Pt multilayer structure. A current flow $I_w$ in Pt layers invoke the spin Hall effect[24, 25] and injects a spin current with a spin polarization $I_s$ into the NiO layer by $I_s = (\hbar/2e)\theta_s I_w \times p$ where $\hbar$ is the reduced planck constant, $e$ is the elementary charge, $\theta_s$ is the spin Hall angle, and $p$ is the unit vector parallel to the flow of the spin current. The spin current with the spin polarization $I_{s,1}$ from the top Pt layer exerts a spin torque $\tau_1$ on the magnetic moment $m_1$ in the NiO layer as $\tau_1 \propto m_1 \times I_{s,1} \times m_1$. The spin current from the bottom Pt layer similarly results in a spin torque $\tau_2 \propto m_2 \times I_{s,2} \times m_2$. Assuming the antiferromagnetic order is coherent in the thickness direction and also assuming a situation where two sets of the spin torque act on the bipartite magnetic moments in the same rotation direction as depicted in Fig. 1 (a), the magnetic moments can efficiently rotate without a cost to increasing the exchange energy (or to competing with the molecular field) which ties the neighboring moments. The magnetic moments rotate until they become orthogonal to the current flow and no more spin torque is exerted. In order to demonstrate the control of the antiferromagnetic moments, we used the Hall bar structure with an experimental procedure, described in Fig. 1 (b). A writing current $I_w$ flowing from the electrode 2 and 3 to the electrodes 1 and 4, as represented by write "1", rotates the magnetic moments and stabilizes them orthogonal to the direction of $I_w$. In the same manner, another writing current $I_w$ flowing from the electrode 2 and 4 to the electrode 1 and 3 writes "0". The orientation of the magnetic moments is read, after each write, by the transverse resistance ($R_{Hall}$) with a small excitation current $I_r$ flowing from the electrode 1 to 2. We particularly took advantage of the spin Hall magnetoresistance to read out the orientation of the magnetic moments. In the phenomenology of the spin Hall magnetoresistance[4], the longitudinal and transverse resistance vary by the relative angle between the magnetic moments in a magnetic layer and the spin polarization of the spin current created by the spin Hall effect in a non-magnetic layer. By adopting the equation for ferromagnetic case[4] to our geometry, $R_{Hall}$ due to the orientation of the NiO magnetic moments can be written as,

$$R_{Hall} \propto -\Delta R_{SMR} \sin\theta_n \cos\theta_n, \tag{1}$$

where $\Delta R_{SMR}$ is the coefficient of the spin Hall magnetoresistance and $\theta_n$ is the angle of the Néel vector $\boldsymbol{n}$ with $\boldsymbol{n} = \boldsymbol{m_1} - \boldsymbol{m_2}$ (see Fig. 1(b) for our circular coordinate with the definition of the general angle $\theta$.). $\boldsymbol{m_1}$ and $\boldsymbol{m_2}$ represent a unit vector for the bipartite magnetic moments as depicted in Fig. 1(a). Here we only consider the in-plane orientation of the magnetic moments.

Figure 1 (c) shows representative results of the sequential write-read operation in Pt/ NiO $t_{NiO}$ nm /Pt and Pt/ SiO$_2$ 10 nm/ Pt with $I_w$ = 38 mA. The operation of write "1" results in a high resistance state and "0" in a low state, which is coherently explained by the spin torque rotation of the magnetic moments and the change of $R_{Hall}$ due to the spin Hall magnetoresistance described in Eq. (1). For instance, with the write "1", the spin torque directs the Neel vector at $\theta_n$ = 135° or 315° (these two possibilities are degenerated in our experiment.) as the magnetic moments $m_1$ and $m_2$ become orthogonal to the direction of $I_w$. Equation (1) yields a high resistance state with $\theta_n$ = 135° or 315° after write "1" and a low state with $\theta_n$ = 45° or 225° after write "0". One can also clearly see the $\Delta R_{Hall}$, which is defined as the change of $R_{Hall}$ after each writing operation, varies with $t_{NiO}$. The control experiments in Pt/ SiO$_2$/ Pt, where SiO$_2$ is a diamagnetic material, show no $\Delta R_{Hall}$. We note that the results of the write-read operations are found irrelevant to the polarity of $I_w$.

Figure 2 (a) shows $I_w$ dependence of $\Delta R_{Hall}$ for Pt/ NiO 20 nm/ Pt. We observed a threshold around 30 mA, corresponding to the current density of 5 x 10$^7$ A/cm$^2$. $t_{NiO}$ dependence of $\Delta R_{Hall}$ shown in Fig. 2 (b) peaks around 10 nm and monotonously decreases with increasing $t_{NiO}$. Appreciable $\Delta R_{Hall}$ was observed up to $t_{NiO}$ = 100 nm. We confirmed that the NiO is unsusceptible to an external magnetic field up to 9 Tesla as Fig. 2 (c) shows the representative data for Pt/ NiO 10nm/ Pt.

In order for our spin torque writing scheme (Fig. 1 (a)) to work, the magnetic moments of the NiO need to lie in the sample plane and the staggered magnetic moments are coherent in the thickness direction. To address this point, we investigated the crystal orientation of the Pt/NiO/Pt layer stacks by X-ray diffraction (XRD). Figure 3 (a) shows the XRD spectra for representative Pt/NiO/Pt layer stacks. We found diffraction peaks only from (111) planes of the Pt layer as well as of the NiO layer, and found no peaks from the (200) planes, which essentially dictates that the crystalline texture of both the Pt and the NiO are predominantly facing (111), on which the magnetic moments lie, on the sample plane. It should be noted that the observed NiO (111) peaks were found at lower angle than that expected in the bulk [26], indicating that the NiO layer sustains a compressive stress[27]. The intensity of the NiO (111) peak, representing degree of the (111) texturing, rapidly develops below $t_{NiO} \sim 10$ nm. The rapid development of the peak intensity coincides with the onset of $\Delta R_{Hall}$ seen in Fig. 2 (b), suggesting that the alignment of magnetic moments is critical for our writing scheme. The $\Delta R_{Hall}$ trend above $t_{NiO} \sim 10$ nm may suggest that the efficiency of the present spin torque writing scheme depends on the total volume of the antiferromagnet in the similar fashion to the case of ferromagnets as $\Delta R_{Hall}$ indeed linearly changes with $1/t_{NiO}$. The activation energy to overcome in the present case should be a total magnetic anisotropy energy being proportional to the volume of the antiferromagnet. One may also notices that our writing scheme requires the NiO layer to have oppositely pointing magnetic moments at the top and bottom interfaces as exactly depicted in Fig. 1 (a) (Otherwise, two sets of the spin torque compete with each other.). We believe that a half of the device area is suitable for the writing scheme as the chance of being such a magnetic configuration should be 50 %. It should also be noted that we observed a memristive behavior in Pt/ NiO/ Pt (see Supplementary information), implying that the spin torque rotation of the magnetic moment occurs portion by portion, possibly as domains, in the NiO. Since domain nucleation and propagation are commonly considered in the spin orbit torque switching in ferromagnets [28], a similar behavior should be expected in the antiferromagnetic cases. As, however, the antiferromagnetic magnetic domain structures are generally quite complex [29] unlike those of ferromagnets, detailed understanding of the antiferromagnetic domain motions under the spin toque requires further extensive investigations.

In summary, we successfully demonstrated the spin-torque control and the resistive read of the magnetic moments in NiO which is an archetype collinear antiferromagnet. The scheme of controlling and detecting antiferromagnets presented here can be applicable to a wide variety of collinear antiferromagnets, and perhaps other types of antiferromagnets. We emphasize that the demonstrated spin-torque control of NiO is apparently much more efficient than a field control requiring > 9 Tesla. The detection scheme using the spin Hall magnetoresistance is much more easily accessible than traditionally used neutron scattering[20] as well as X-ray magnetic linear dichroism [30]. Thus, those of our easily accessible methodologies will open up more fundamental research opportunities on antiferromagnets. Ultimately, we stress that basic requirements for practical antiferromagnetic spintronic devices, i.e. the electrical control and detection of antiferromagnetic moments, are now fulfilled.

**Acknowledgements**

We would like to thank Prof. Koki Takanashi and Prof. Takeshi Seki for valuable discussions and for comments on the manuscript. We also thank Prof. Daisuke Kan for helping us with performing the X-ray diffraction measurements. This work is supported by JSPS KAKENHI Grant Numbers 15H05702, 26870300, 17H04924, 17H05181. We also acknowledge The Cooperative Research Project Program of the Research Institute of Electrical Communication, Tohoku University.

**Author contributions**

T. M. and T. O. conceived the idea. T. M. planned the experiments. T. M. and K.O. conducted the measurements, and collected and analyzed the data. T. M. wrote the manuscript with input from T. O. All the authors discussed the results.

**Additional Information**

Correspondence and requests for materials should be addressed to T. M.

**Competing financial interests**

The authors declare no competing financial interests.

**Methods**

**Sample fabrication and characterization.** The multilayers were formed by magnetron sputtering on a thermally oxidized Si wafer. We used a sintered NiO sputtering target for the deposition of the NiO layer. All the deposition was done at room temperature with a base pressure below $2 \times 10^{-6}$ Pa. The X-ray diffraction with a Cu Kα radiation was performed in the blanket films of the multilayer. The Hall bar structure with 5 μm-wide channel was patterned by a conventional photolithography followed by an Ar ion milling. Ti (5nm)/Au (100nm) was deposited for the contact pads. All the measurements were performed at room temperature.

**The spin torque write-read procedure.** We used a multiplexer system built with Keithley 2700 to quickly route the current flow and voltage probes among the electrodes 1~4. We applied $I_w$ for 3 seconds to write, then shut it off, and started reading $R_{Hall}$ with an excitation current $I_r$ of 1 mA. $R_{Hall}$ was read multiple times with an interval of 30 seconds.

Figure captions:

**Figure 1 The spin torque writing scheme and the sequential write-read memory operation.** (a) The basic principle of the spin torque rotation of the antiferromagnets using the spin Hall effect in a Pt/NiO/Pt multilayer structure. The writing current $I_w$ in the Pt layers invokes the spin Hall effect, and injects the spin currents with the spin polarization $I_{s,1}$ and $I_{s,2}$ toward the antiferromagnetic NiO. The spin currents exert a spin torque $\tau_1$ and $\tau_2$ (green arrows) on the microscopic magnetic moments in the same rotation direction. (b) Measurement procedure of the spin torque write and the Hall resistance read. $R_{Hall}$ is measured after each write "1" and "0". Microscope image of the Hall cross structure is shown in the center. Our common definition of the angle $\theta$ is also shown. (c) The sequential write-read operation in Pt/ NiO $t_{NiO}$ nm /Pt and Pt/ SiO$_2$ 10 nm/ Pt with $I_w$ = 38 mA. The arrows on the top represent the write "1" and "0" operations described in (b). As indicated in the graph, $\Delta R_{Hall}$ is defined as $\Delta R_{Hall}$ = $R_{Hall}$ ("1") - $R_{Hall}$ ("0") where $R_{Hall}$ ("1") and $R_{Hall}$ ("0") respectively represent the $R_{Hall}$ after the write "1" and the write "0".

**Figure 2 Writability by the spin torque and robustness against an external field.** (a) $\Delta R_{Hall}$ as a function of the writing current $I_w$ for Pt/ NiO 20 nm /Pt. (b) $\Delta R_{Hall}$ as a function of the thickness of the NiO and the SiO$_2$. The inset shows $\Delta R_{Hall}$ as a function of the reciprocal thickness of the NiO. The dotted line is the linear fitting in the thickness range of $t_{NiO}$ = 20 ~ 100 nm. (c) $R_{Hall}$ measurements for Pt/ NiO 10 nm /Pt in a rotating magnetic field $H_{ex}$ =1, 5, and 9 Tesla. $\theta_H$ is the field angle in the circular coordinate shown in Fig. 1 (b). $R_{Hall}$ curves are offset for visibility.

**Figure 3 X-ray diffraction (XRD) on Pt/NiO/Pt.** (a) XRD spectra for Pt/ NiO $t_{NiO}$ nm/Pt with $t_{NiO}$ = 0, 5, 20, 50, 90 nm. The dotted lines indicate the expected diffraction angles for bulk NiO and Pt referenced from Ref. 30. (b) The NiO(111) peak intensity as a function of $t_{NiO}$.

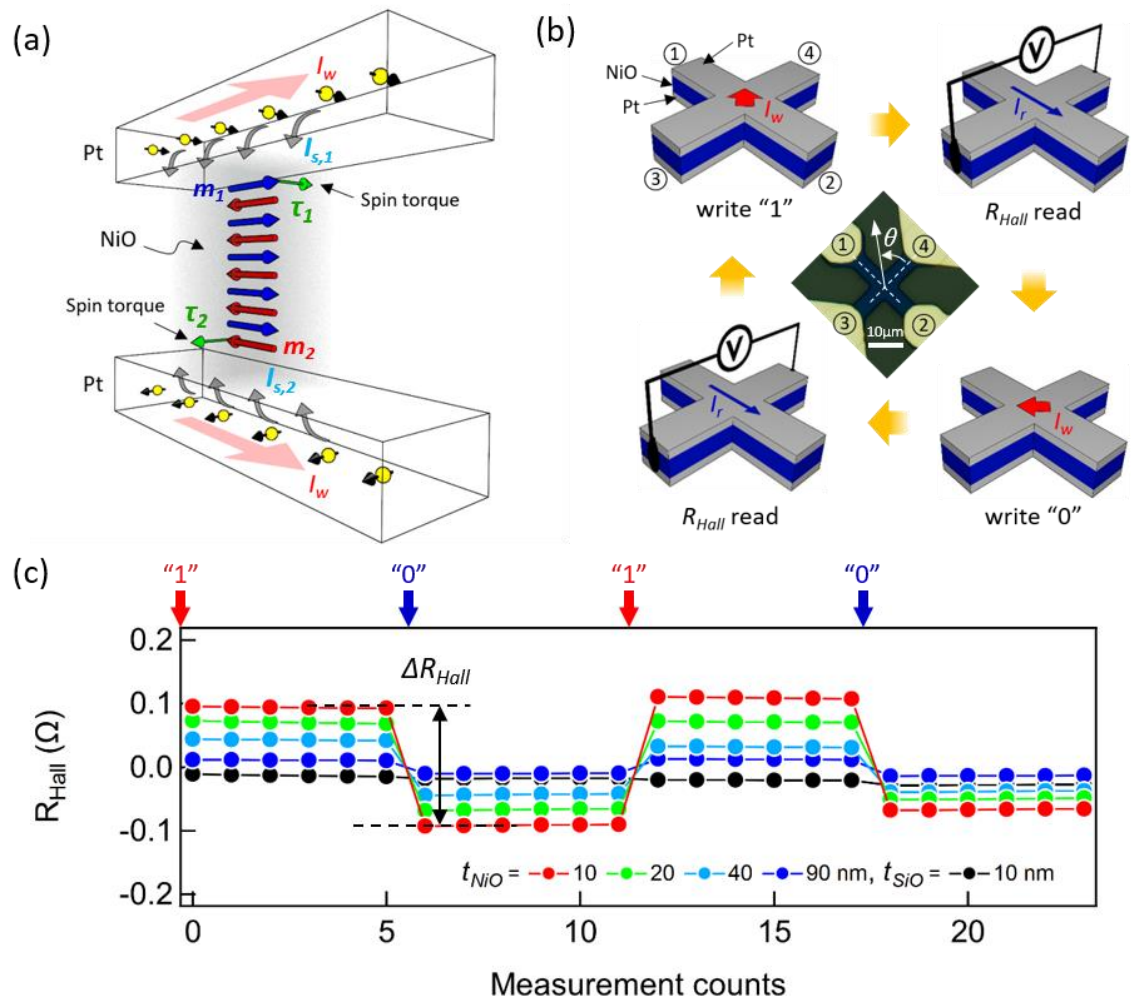

Figure 1 Moriyama et al.

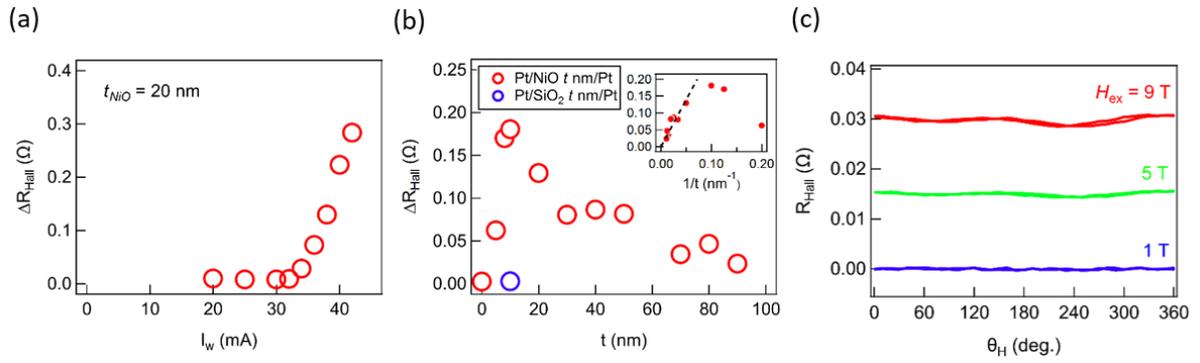

Figure 2 Moriyama et al.

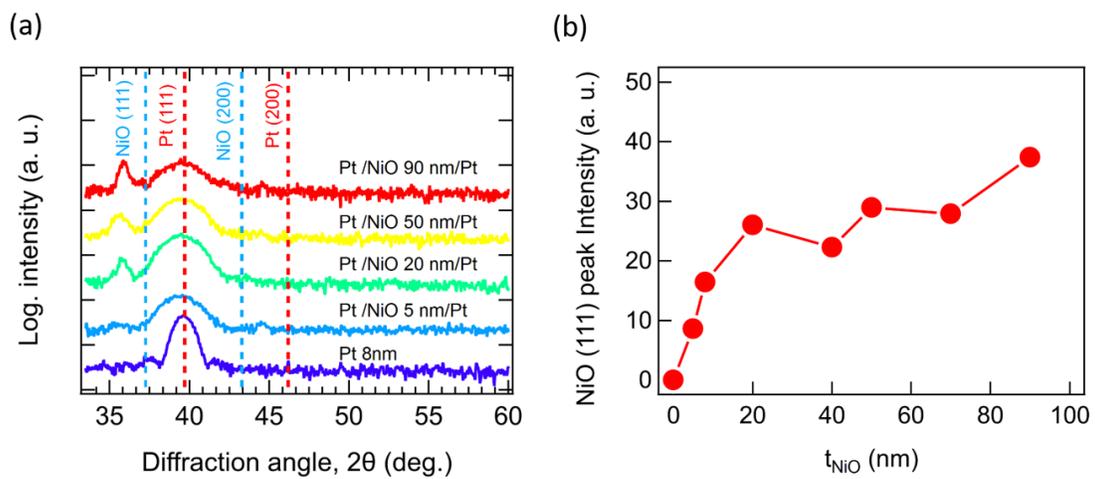

Figure 3 Moriyama et al.